\documentclass[prb,twocolumn]{revtex4}

\usepackage{amsmath}
\usepackage{graphicx}

\begin{document}

%%%%%%%%%%%%
\title{From String Nets to Nonabelions}

\author{Lukasz Fidkowski$^{1,4}$, Michael Freedman$^1$, Chetan Nayak$^{1,2}$,
Kevin Walker$^1$, and Zhenghan Wang$^{1,3}$}
\affiliation{$^1$Microsoft Station Q, University of California, Santa Barbara 93106-6105\\
$^2$ Department of Physics and Astronomy, University of California,
Los Angeles, CA 90095-1547\\
$^3$ Department of Mathematics, Indiana University,
Bloomington, IN 47405\\
$^4$ Department of Physics, Stanford University, Stanford, CA 94305}

\date{\today }

\begin{abstract}
We discuss Hilbert spaces spanned by the set of string nets, i.e. trivalent
graphs, on a lattice. We suggest some routes by which such a Hilbert space
could be the low-energy subspace of a model of quantum spins on a lattice
with short-ranged interactions. We then explain conditions which a
Hamiltonian acting on this string net Hilbert space must satisfy in order
for its ground state and low-lying quasiparticle excitations
to be in the DFib topological phase. Using the string net wavefunction,
we describe the properties of this phase. Our discussion is informed by
mappings of string net wavefunctions to the chromatic polynomial
and the Potts model.
\end{abstract}

\maketitle
%%%%%%%%%%%%
\section{Introduction}

The mathematical theory of anyons - modular tensor categories - is extremely rich in
examples, and existing physical theory provides a way of describing most of these examples
as effective Chern-Simons gauge theories.  In contrast, our knowledge is limited when it
comes to identifying plausible solid state Hamiltonians from which a (2D) state of matter could
emerge whose effective low energy description is, in fact, a Chern-Simons theory.  The
off-diagonal conductivity of fractional quantum Hall (FQHE) systems is tantamount to the
equation of motion for a Chern-Simons Lagrangian, so Hall systems are the most developed
source of such examples.  The best-studied example among abelian states is the $\nu=1/3$
Laughlin state.  The foremost candidate among non-Abelian states is the Pfaffian
state \cite{Moore91}, which is believed \cite{Morf98,Rezayi00} to be realized at the
$\nu=5/2$ fractional quantum hall plateau.
Beyond $\nu=5/2$, more delicate plateaus at $\nu=12/5, 4/7$, etc. may also support
nonabelions.  However, in this paper we explore a quite distinct family of Hamiltonians.

Because magnetic interactions in solids can be at energy scales as high as
$\sim 10^3$ Kelvin, it would be very exciting to find realistic families of spin Hamiltonians
representing a nonabelian phase.  (This has essentially been accomplished\cite{Z2}
for the simplest abelian phase, $Z_2$ gauge theory, although the
corresponding experimental system has not been clearly
identified.)  This goal has been pursued for several years through the study of model
Hamiltonians $H$ acting in an effective Hilbert space $\cal H$ whose degrees of freedom are
either unoriented loops \cite{FNS}, or, more recently, branching loops called
``string nets''  \cite{LWstrnet}.  Such Hilbert spaces $\cal H$ are a kind of half-way house.
Eventually, it will be necessary to understand how local spins can encode effective loops and
nets, and some ideas on encoding nets are presented in Section II.  However, the premise
of this paper is that we already have a Hilbert space $\cal H$ spanned by the simplest type of
string net $G$ - the lines are unoriented and unlabeled, the nodes have valence $3$ and lack
internal states.  Our goal then is to formulate, in the most general terms, what properties a
Hamiltonian $H: \cal{H} \rightarrow \cal{H}$ should have to describe the simplest topological
phase of string nets, the ``doubled Fibonacci theory'', DFib.  DFib is not only nonabelian but
actually computationally universal \cite{FLW} and thus an extremely attractive target phase.
The approach here is intended to complement the ideas presented in Refs.~\onlinecite{FF}
and \onlinecite{LWstrnet}.

To summarize our approach in a phrase: ``nature abhors a degeneracy''.  (Consider, for
example, eigenvalue repulsion for random Hamiltonians).  There is an irony here because
topological phases are nothing else than a degenerate, yet stable, ground state for which no
classical symmetry exists to be broken.  From this viewpoint, we will see that building DFib
(and other phases?) amounts to setting a trap for nature.  By compelling a certain space
$V(D,n)$ of low energy modes for a given fixed boundary condition (specified by integer $n$,
as we describe below) to have dimension equal to $d$, unexpectedly small, we trap a
class of Hamiltonians into an exponential growth of degeneracy:
$\text{lim}_{n\rightarrow \infty} {\left( \text{dim} (V(D,n)) \right)}^{1/n} = \tau =
\frac{1+\sqrt{5}}{2}$.  We present a rather surprising derivation of DFib from dimensional
considerations alone; the $6j$-symbol derives from the assumption of unitarity and minimal
dimension of ``disk spaces".  Although the $6j$-symbol obeys the pentagon equations we do
not use the pentagon equation to find the $6j$-symbol.  The physical significance is that DFib
should be a robust phase stabilized by a type of eigenvalue repulsion.

After DFib is derived, a beautiful formula of Tutte (compare Ref.~\onlinecite{FF}) allows us to
make an exact connection to the $Q=\tau+2 \approx 3.618$ state Potts model, where
$\tau = \frac{1+\sqrt{5}}{2}$, the golden ratio.  We find that
the exactly solvable point in the DFib phase is the high temperature limit of the low
temperature expansion of the ${\tau+2}$-state Potts model.  This allows us to conjecture
that the topological phase extends downward in ``temperature'' until the critical point is reached at $\text{log } \beta = \sqrt{\tau+2} + 1$.
This suggests a one-parameter family of DFib-Hamiltonians whose ground state wave
functions are not strictly topological but have a ``length'' or bond-fugacity, $x$,
satisfying $0.345 \approx {1}/(\sqrt{\tau+2}+1) \leq x \leq 1$, implying a
considerable stability within this phase.

The paper is organized as follows. In Section II, some ideas are presented
for how string nets could emerge from microscopic models of quantum spins on
a lattice. In Section III, DFib is derived from dimensional
considerations.  In Section IV, we describe, in string net language,
the quasiparticle excitations of DFib. In Section V, string net wavefunctions
and their squares are related to the chromatic polynomial.  In Section VI we show that the
topological string net wave function has a ``plasma analogy'' to the $\tau+2$-state Potts model,
and in Section VII we discuss our conclusions. In the appendix, Baxter's hard hexagon model is
used to extend a theorem of Tutte's.

\section{How to construct a net Hilbert space $\cal H$}

\begin{figure}[tbh!]
\includegraphics[width=3.25in]{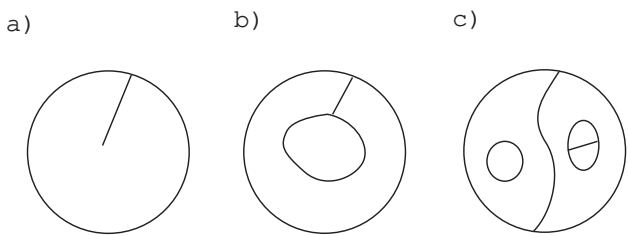}
\caption{a) is not in the Hilbert space ${\cal H}$, but b) and c) are.
In these figures, the outermost circle is the boundary of the system, where
nets are allowed to terminate. The endpoint in the middle in (a) is a violation
of the ``no dead ends'' condition.}
\label{fig:psnf2}
\end{figure}

In this paper, we will be concerned with Hamiltonians $H$ acting on Hilbert
spaces ${\cal H}(\Sigma)$ of wave functions $\Psi$ that assign complex-valued amplitudes
to string nets (``nets'' for short) on a surface $\Sigma$.
The surfaces $\Sigma$ which could be relevant to experimental systems
are presumably planes with some number of punctures. However,
it is quite profitable conceptually and for the purpose of numerical simulations
to think about higher-genus surfaces as well.
A net is what mathematicians
call a trivalent graph; it has only simple branching and no ``dead ends" (univalent vertices)
except as defined by boundary conditions at the edge of the surface.
According to this definition, (the Dirac function on the net in) Figure \ref{fig:psnf2}
a) is not in the Hilbert space $\cal H$ but (those of) Figure \ref{fig:psnf2} b) and \ref{fig:psnf2} c) are.
In most of this paper, we will simply assume that the
Hilbert spaces ${\cal H}(\Sigma)$ arise as the low-energy subspaces
of the Hilbert spaces of a system of spins or electrons in a solid or ultra-cold
atoms on an optical lattice. In such a formulation Figure \ref{fig:psnf2} a) could be
thought of as a high, but finite, energy state of the spins, electrons in a solid or ultra-cold
atoms.

In this section, however, we will consider
the question: from what kinds of lattice models do nets emerge
in the low-energy description so that the
Hilbert spaces ${\cal H}(\Sigma)$ are the low-energy subspaces?
Three ideas A, B, and C are sketched for writing a spin Hamiltonian
$K: \bar{H} \rightarrow \bar{H}$ on a large Hilbert space $\bar{H}$ of
microscopic degrees of freedom so that the ground state manifold $\cal H$ of $K$
will be the ``Hilbert space of nets" on which this paper is predicated.

All our $K$ break $SU(2)$-invariance and require fine tuning.
Ideas A and B are conceptually very simple but both require a three-body interaction.
Idea C is really an encryption of B into a 2-body interaction on higher spin (spin $= 3/2$) particles.

\vspace{6mm}
{\bf A.}
\vspace{2mm}

$\bar{H} = \otimes_{\text{bonds}} C^2$, i.e. is a Hilbert space of spin $= 1/2$ particles
living on the links of a trivalent graph such as the honeycomb.  We interpret an ${S_i^z}=1/2$
link as one on which the net lies. We take $K = \Sigma_{\text{sites}} K_s$ where
$K_s$ projects onto the subspace of $C^2 \otimes C^2 \otimes C^2$ (i.e. of the Hilbert
space of the three spins surrounding a lattice) of total  ${\sum_{i=1}^3}{S_i^z}= -1/2$. 
This forbids dead ends. Unfortunately, the function
\begin{eqnarray}
\text{total $S^z$ eigenvalue} &\rightarrow& \text{energy} \cr
\vspace{4mm}
-3/2&\rightarrow&0, \cr
-1/2&\rightarrow&\text{nonzero} \cr
1/2&\rightarrow&0 \cr
3/2&\rightarrow&0
\end{eqnarray}
is not quadratic (no parabola passes through $(-3/2,0), (-1/2, \text{nonzero}), (1/2,0)$, and $(3/2,0)$.
Hence, when $K_s$ is expanded in products of ${\sigma^z}_i$, $i=1,2,3$, running over the bonds
meeting $s$, it must contain a cubic ${\sigma^z}_1 {\sigma^z}_2 {\sigma^z}_3$ term.

\vspace{6mm}
{\bf B.}
\vspace{2mm}

\begin{figure}[tbh!]
\includegraphics[width=1in]{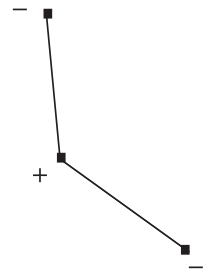}
\caption{An illegal pair of bonds which is energetically penalized by $K_p$.}
\label{fig:psnfa1}
\end{figure}

\begin{figure}[tbh!]
\includegraphics[width=1in]{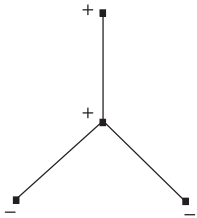}
\caption{An up-spin which has only a single up-spin neighbor. This is
energetically penalized by $K = \Sigma_p K_p$.}
\label{fig:psnfa2}
\end{figure}

An alternative is to put a spin $= 1/2$ particle at the sites of the honeycomb, so
${\cal H} = \otimes_{\text{sites}} C^2$, $C^2 = \langle +, - \rangle$.  For each
consecutive pair $p$ of bonds, $K$ has a term $K_p$: $K = \Sigma_p K_p$,
where $K_p$ is a diagonal matrix all of whose entries are $0$ except for that corresponding to the illegal pair of edges shown in fig.~\ref{fig:psnfa1}).
$K$ penalizes both isolated $+$'s and $+$'s with exactly one $+$ neighbor. The latter
situation is shown in fig.~\ref{fig:psnfa2}.
Interpreting those bonds bounded by two $+$ signs as the ones on which the net lies,
we see that the zero modes of $K$ are precisely the nets (trivalent graphs) within the honeycomb.  Unfortunately $K$ seems resolutely $3$-body.

\vspace{6mm}
{\bf B'.}
\vspace{2mm}

\begin{figure}[tbh!]
\includegraphics[width=2in]{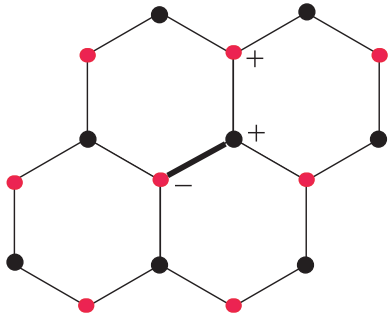}
\caption{Upon reversing the spins on one sublattice of the honeycomb lattice,
$\tilde{K}_p$ and $\tilde{K'}_{p'}$ now penalize maximum and minimum $S^z$
eigenvalues respectively.}
\label{fig:psnfa4}
\end{figure}

To set the stage for our final construction, it is helpful to reverse $+$ and $-$ spins
on the index $2$ Bravais lattice $L'$ within the honeycomb, honeycomb $ = L \bigcup L'$.  With this convention, $\tilde{K} = \sum_{\text{p centered on L}} \tilde{K}_p +
\sum_{\text{p' centered on L'}} \tilde{K'}_{p'}$ where $\tilde{K}_p$ projects to the
highest total $S_z$ eigenvalue, ${S_z}= 3/2$,
and $\tilde{K'}_{p'}$ projects to the lowest total $S_z$ eigenvalue, ${S_z}=-3/2$.
In other words, the penalized configurations are 3 consecutive pluses or 3 consecutive minuses.
 Also, we now draw bonds between plus sites of $L$ and minus sites of $L'$ ($L$ and $L'$ labeled as black and red respectively in Figure \ref{fig:psnfa4}).    $\tilde{K}$ is still necessarily $3$-body but at least it now has the form of the Klein Hamiltonian, see e.g. ref.~\onlinecite{CCK}

\vspace{6mm}
{\bf C.}
\vspace{2mm}

\begin{figure}[tbh]
\includegraphics[width=2.5in]{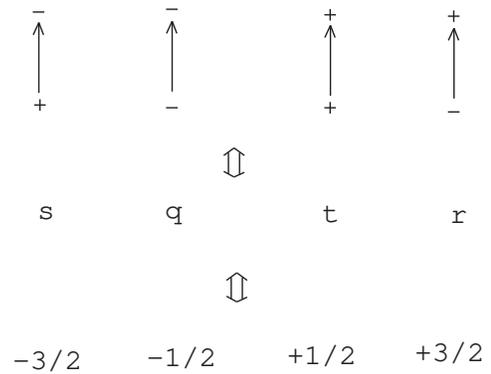}
\caption{the bottom row represents the spin eigenvalues}
\label{fig:psnfa3}
\end{figure}

The idea here is to take adjacent pairs of site spins from $B'$ and encrypt them as the state of a spin $= 3/2$ particle living on the bond $b$ joining the two sites.  We orient $b$ from $L'$ to $L$.  Let us set up an indeterminate bijection (fig.~\ref{fig:psnfa3}).
We now ``simulate" $\tilde{K}$ on a Hilbert space $\bar{\cal H} = \otimes_{\text{bonds}} C^4$, the space of a spin $= 3/2$ particle on each bond.  On sites $l \in L$ ($l' \in L'$) we must penalize $t \otimes t$ ($q \otimes q$).

Furthermore we must penalize inconsistent encryptions.  For $l \in L$ ($l' \in L'$) the following pairs yield inconsistent site labels: $q \otimes r, r \otimes q, q \otimes t, t \otimes q, s \otimes r, r \otimes s, s \otimes t$, and $t \otimes s$ ($q \otimes s, s \otimes q, q \otimes t, t \otimes q, r \otimes s, s \otimes r, r \otimes t$, and $t \otimes r$).

We again obtain the space of nets as zero modes ${\cal H} \subset \bar{\cal H}$ by fixing $K$ to be the following $2$-body Hamiltonian:
\begin{eqnarray}
K&=&\sum_{l \in L} \text{(} \Pi_{t \otimes t} + \Pi_{q \otimes r} + \Pi_{r \otimes q} + \Pi_{q \otimes t} +\Pi_{t \otimes q} \nonumber \\ &+& \Pi_{s \otimes r} +\Pi_{r \otimes s} + \Pi_{s \otimes t} +\Pi_{t \otimes s} \nonumber \text{)}\\
&+&\sum_{l' \in L'} \text{(} \Pi_{q \otimes q} +\Pi_{q \otimes s} +\Pi_{s \otimes q} +\Pi_{q \otimes t} +\Pi_{t \otimes q} \nonumber \\ &+&\Pi_{r \otimes s} +\Pi_{s \otimes r} +\Pi_{r \otimes t} +\Pi_{t \otimes r} \text{)},
\end{eqnarray}
where $\Pi_{x \otimes y}$ denotes the projector onto the state $x \otimes y$.

\begin{figure}[tbh!]
\includegraphics[width=2.5in]{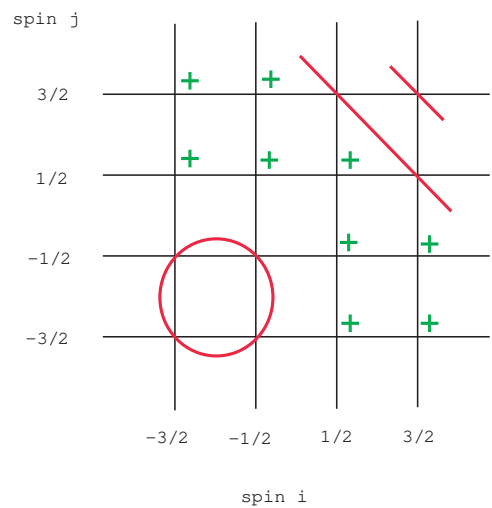}
\caption{The horizontal axis labels the $i$th spin and the vertical axis the $j$th spin.}
\label{fig:psnfa5}
\end{figure}

We can write a Hamiltonian which effectively accomplishes such a projection in terms of the spins $S^z_i$ ($i$ is the bond index).  For black sites $H$ has terms
\begin{eqnarray} \left( \left( S^z_i + 1 \right)^2 + \left(S^z_j + 1 \right)^2 - 1/2 \right) \nonumber \\ \left(S^z_i + S^z_j - 2 \right) \left( S^z_i + {S_z}_j - 3 \right) \end{eqnarray}
and for red sites
\begin{eqnarray} \left( \left( S^z_i - 1 \right)^2 + \left(S^z_j - 1 \right)^2 - 1/2 \right) \nonumber \\ \left(S^z_i + S^z_j + 2 \right) \left( S^z_i + S^z_j + 3 \right). \end{eqnarray}
The origin of these terms is illustrated in Figure \ref{fig:psnfa5}.

\section{$\cal{H}$, $H$, and $V$}

From now on, we will be concerned with he properties of the Hilbert spaces
${\cal H}(\Sigma)$ on a surface $\Sigma$.
If $\Sigma$ has boundary, $\partial \Sigma$, we fix a boundary condition by
specifying points where the nets must end. 
In the case where $\Sigma$ has connected boundary and the boundary condition
consists of $n$ points, we denote the Hilbert space by ${\cal H}(\Sigma,n)$.
We will be interested
in ``isotopy invariant'' wave functions $\Psi$, whose value is independent of deformation.
\begin{figure}[tbh!]
\includegraphics[width=1.5in]{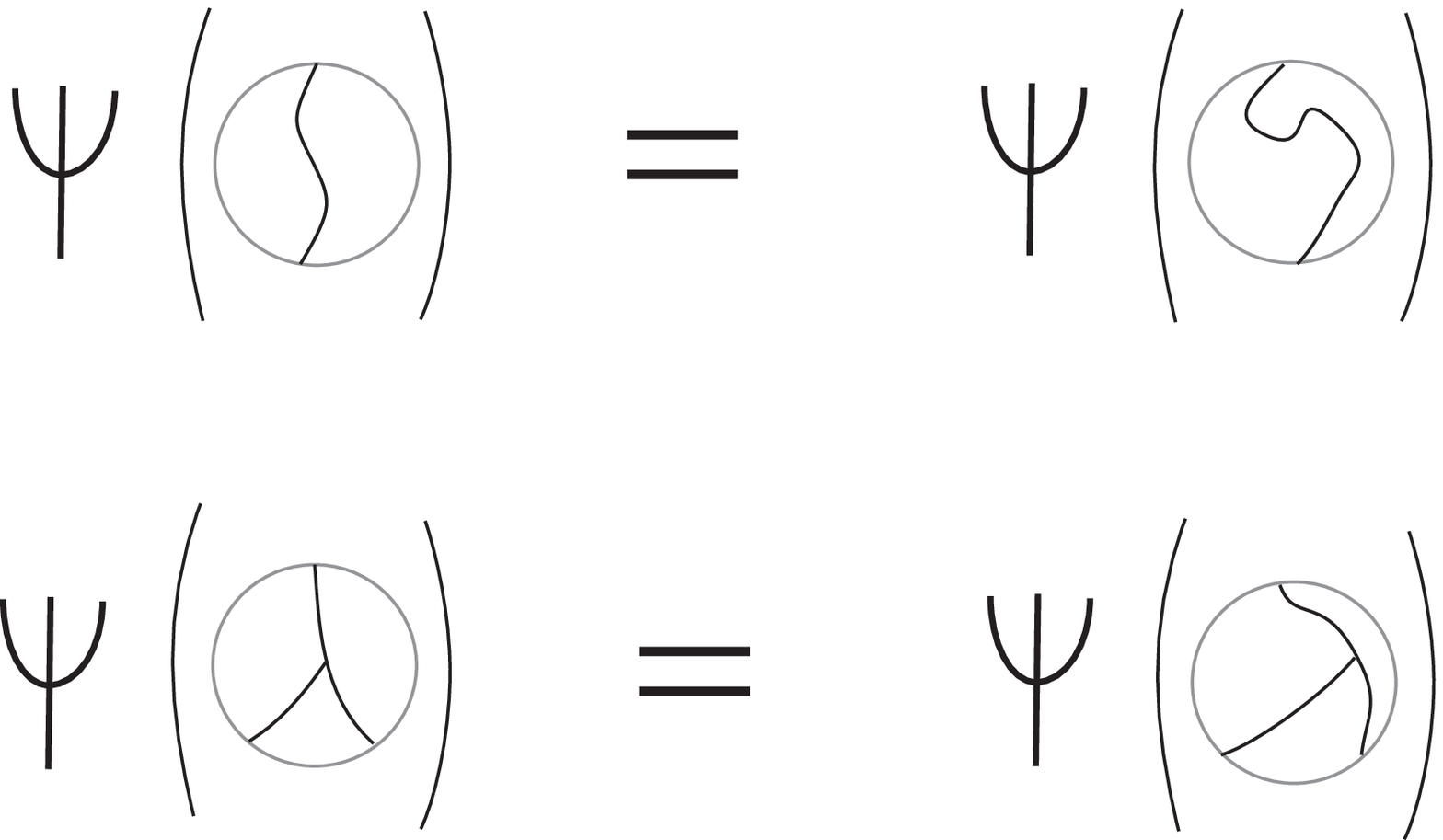}
\caption{The condition of isotopy invariance. Wavefunctions assign a
complex amplitude for any string net (which is the amplitude for this
configuration to occur). These equations mean that
wavefunctions in the low-energy Hilbert space assign the
same amplitude to two string nets if one can be obtained from the
other by smooth deformations.}
\label{fig:psnf1}
\end{figure}
(In Section VI we relax this condition to allow certain bond fugacities.) 
To avoid unnormalizable wave functions, these loops should really live on a lattice,
as in the previous section.

To produce invariant $\Psi$'s we consider Hamiltonians $H: {\cal H} \rightarrow {\cal H}$
which contain fluctuations sufficient to enforce isotopy invariance (fig.~\ref{fig:psnf1}) on all
low energy states.  It should be remarked that it is not easy to set up such terms on a lattice;
some fine tuning may be required.  Also there are questions of ergodicity - $H$ must have
sufficient fluctuations that crystals do not compete with the liquid condition described by
Fig.~\ref{fig:psnf1}.  Nevertheless we start by assuming these problems solved: that we
have $\cal{H}$ and a family $\{ H \}$ whose ground states $V$ consist of isotopy invariant
wave functions $\Psi$.

The ``axioms'' we impose on the Hamiltonian $H$ are implicit in the
following conditions that we require its ground state manifolds
$V(\Sigma,n) \subset {\cal H}(\Sigma,n)$ to satisfy.

{\bf Axiom 1}. $H$ is gapped - this makes $V(\Sigma,n)$ sharply defined.

{\bf Axiom 2.}  The following ``minimal'' dimensions on the 2-disk $\Sigma=D$ occur:

(i) dim $V(D,0)=1$

(ii) dim $V(D,1)=0$

(iii) dim $V(D,2)=1$

(iv) dim $V(D,3)=1$

(v) dim $V(D,4)$ takes the minimal value consistent with (i)-(iv) (to be computed below).

\begin{figure}[tbh!]
\includegraphics[width=2in]{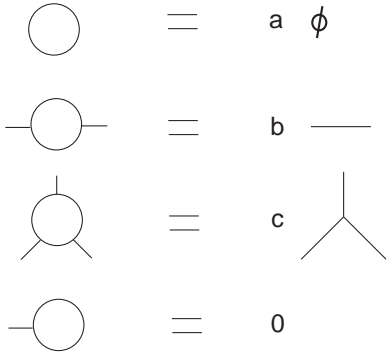}
\caption{We have stopped drawing the disk, but all diagrams above are nets in a $2$-disk $D$ with the endpoints on $\d D$.  Also we abuse notation to allow a net to also represent the ground state evaluation, $\Psi_{\text{g.s.}} (\text{net})$. }
\label{fig:psnf3}
\end{figure}

We make the further technical assumption that the constants $a$, $b$ and $c$
in Figure \ref{fig:psnf3} are neither 0 nor infinity.

Axiom 2(ii) is the ``no tadpole'' axiom which says that although figure
\ref{fig:psnf2} b) is in ${\cal H}$ it has high energy.  There is no low energy
manifold $V$ whatsoever when the boundary condition only allows tadpoles. 
If one thinks in terms of $1+1$ dimensional physics, 2(ii) merely says the obvious:
a single particle should not come out of the vacuum.
If we nevertheless persisted in making dim $(V(D,1))=1$ we would admit the very boring case
in which for all $\Sigma$, dim$(V(\Sigma))=1$ and $V$ is spanned by the constant function
on nets.

Similarly 2(i) and 2(iii) are required in a $1+1$ dimensional (unitary tensor category) context.  Also, if either dimension is $0$ then all $V$ have dimension $0$, by gluing formulae, so the entire theory collapses.

Axiom 2(iv) does represent a choice.  If we instead said dim $(V(D,3))=0$
we would forbid our nets to branch.  Here we know $Z_2$ gauge theory
(i.e.~the toric code) and the doubled semion theory --- both abelian --- can arise.
Possibly higher doubled ${SU(2)}_k$ Chern-Simons theories might also arise
from loop models, but entropy arguments \cite{Levin05}
show that their ground state wave functions cannot be a simple
Gibbs factor per loop \cite{Levin05}.  So 2(iii) is not inevitable but represents
our decision to set up whatever microscopics are necessary to build $({\cal H},
H)$ with branched nets occuring in $V$, i.e. to build a string net model.  We now derive:

\newtheorem{thm}{Theorem}
\begin{thm}
There is a unique theory $V(\Sigma)$ compatible with $1$ and $2$ above.  It satisfies dim $(V(D,4)) = 2$ and is DFib, the doubled fibonacci category.
\end{thm}
{\it Proof:}  By ``Axiom 2'' there are nonzero constants $a,b,c \in {\bf C}$ such that
the conditions in Figure \ref{fig:psnf3} hold.

\begin{figure}[bth!]
\includegraphics[width=2.75in]{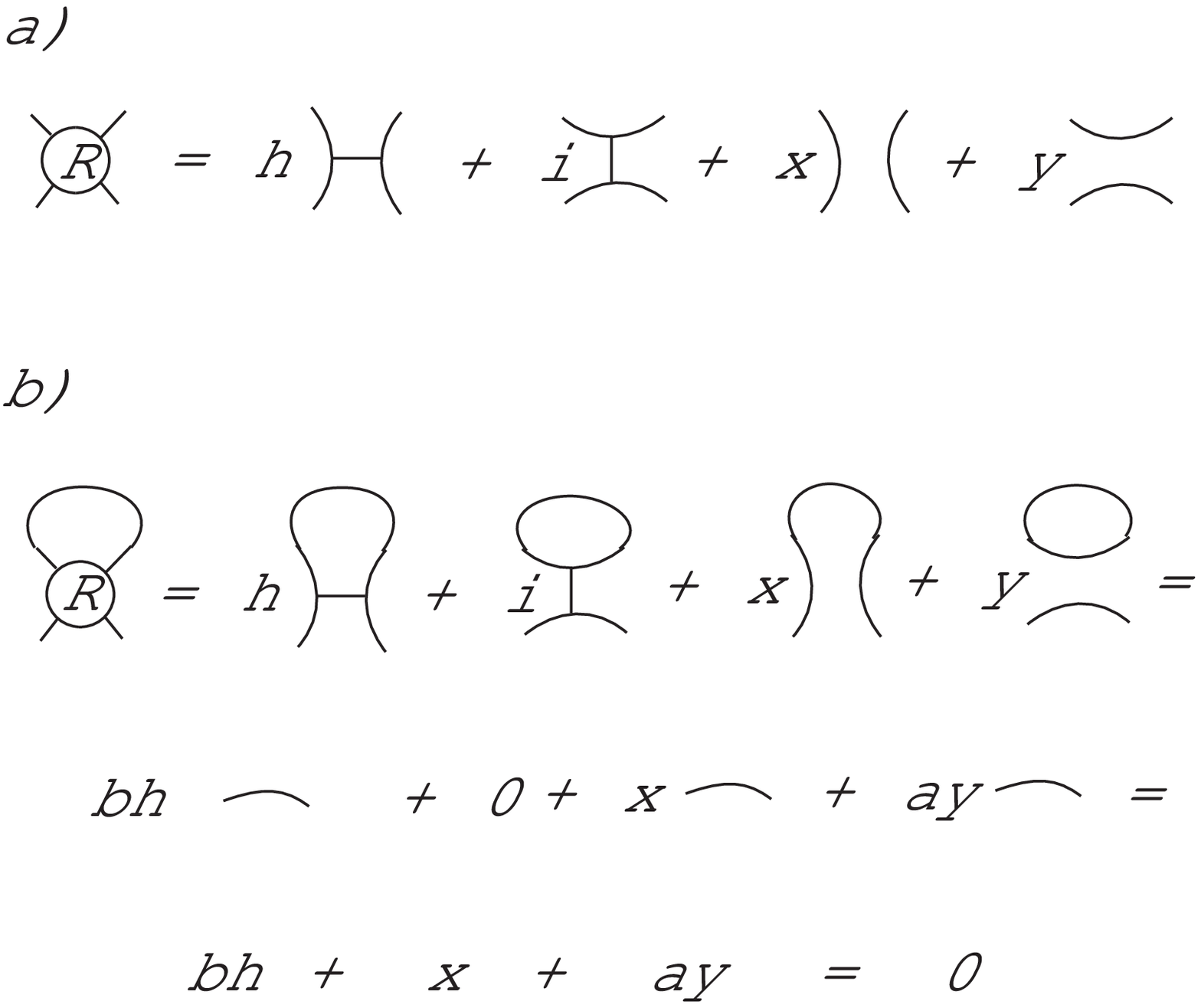}
\caption{If $V(D,4)<4$, then there must be some coefficients $h,i,x,y$
such that the linear combination on the right-hand-side of (a) vanishes.
By embedding these pictures within larger ones in such a way that
the endpoints are connected as shown in (b) and using fig.~\ref{fig:psnf3},
we find relations satisfied by $h,i,x,y$.}
\label{fig:psnf4}
\end{figure}

\begin{figure}[tbh!]
\includegraphics[width=2.75in]{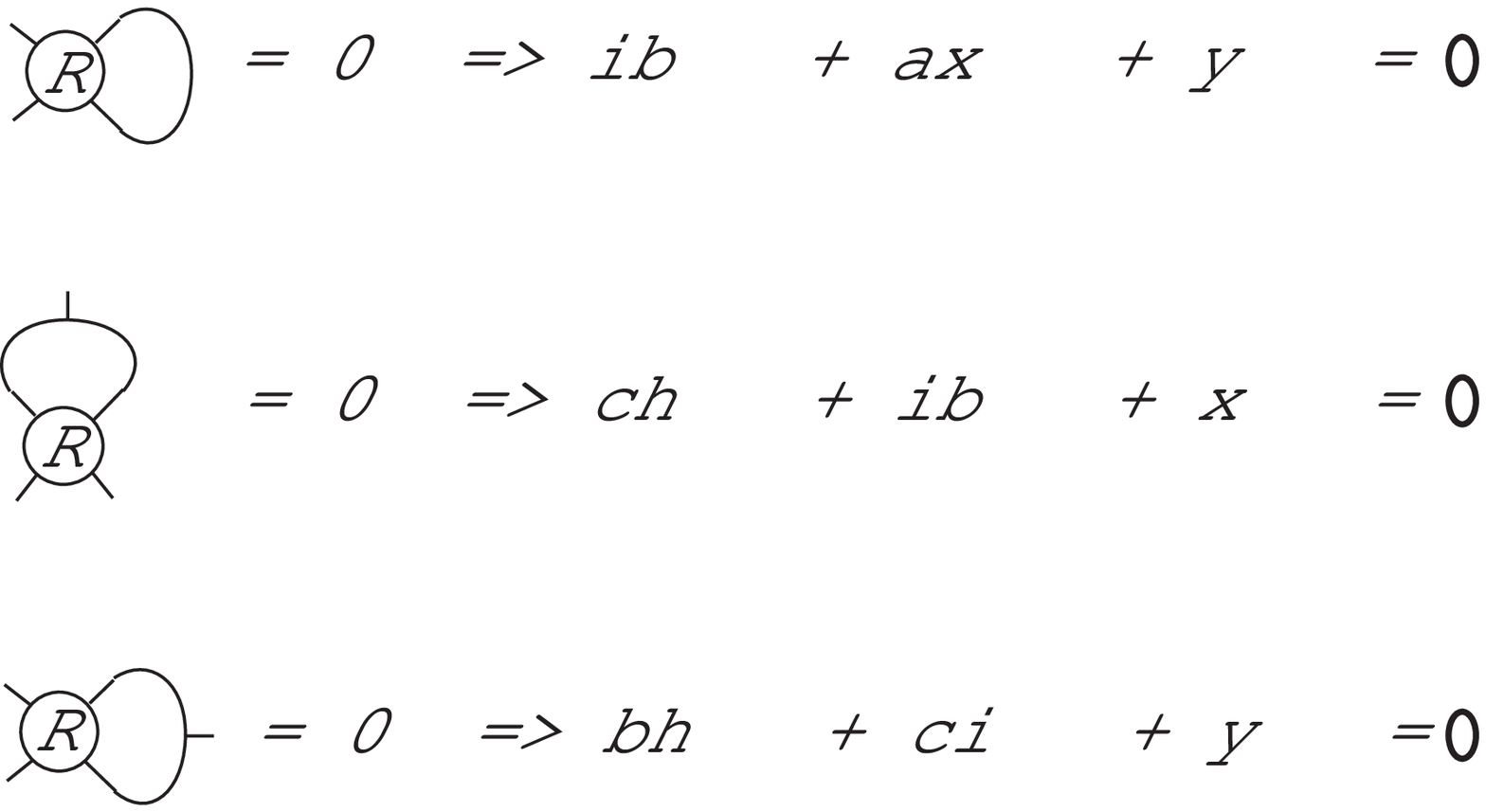}
\caption{Three other relations obtained by connecting the endpoints
of $R$.}
\label{fig:psnf4b}
\end{figure}

Now consider the $4$-point space $V(D,4)$.  If its dimension is $<4$ there must be relations among the $4$ nets in the expression below (fig.~\ref{fig:psnf4} a)), each thought of as an evaluation of the functional $\Psi_{\text{g.s.}}$ on $V(D,4)$.
We find conditions on the coefficients by joining various outputs of R.
This amounts to calculating consequent relations in the $1$-dimensional
spaces $V(D,2)$ and $V(D,3)$ which are implied by $R$. 
We work out, in part b) of fig.~\ref{fig:psnf4} the implication of
joining the upper outputs of $R$ by an arc.  By such arguments,
we also have the three relations in fig.~(\ref{fig:psnf4b}).
Eliminating $x$ and $y$, $x=-ch-bi,y=-bh-ci$, we obtain 
\begin{eqnarray} (b-c-ab)h+(-b-ac)i&=&0 \cr
(-b-ac)i+(b-c-ab)h&=&0. \end{eqnarray}
Possible relations which can be imposed on string nets through
the Hamiltonian correspond to solutions of this linear system.
There can be at most two linearly independent solutions, and this case occurs exactly when the coefficients vanish: $b=c+ab, b=-ac$ so $-ac=c+(a^2)c$ or $a^2=a+1$.  We already see the golden ratio $\tau$ emerge: $a=\tau$ or $-\tau^{-1}$, $\tau=\frac{1+\sqrt{5}}{2}$.

\begin{figure}[tbh!]
\includegraphics[width=2.5in]{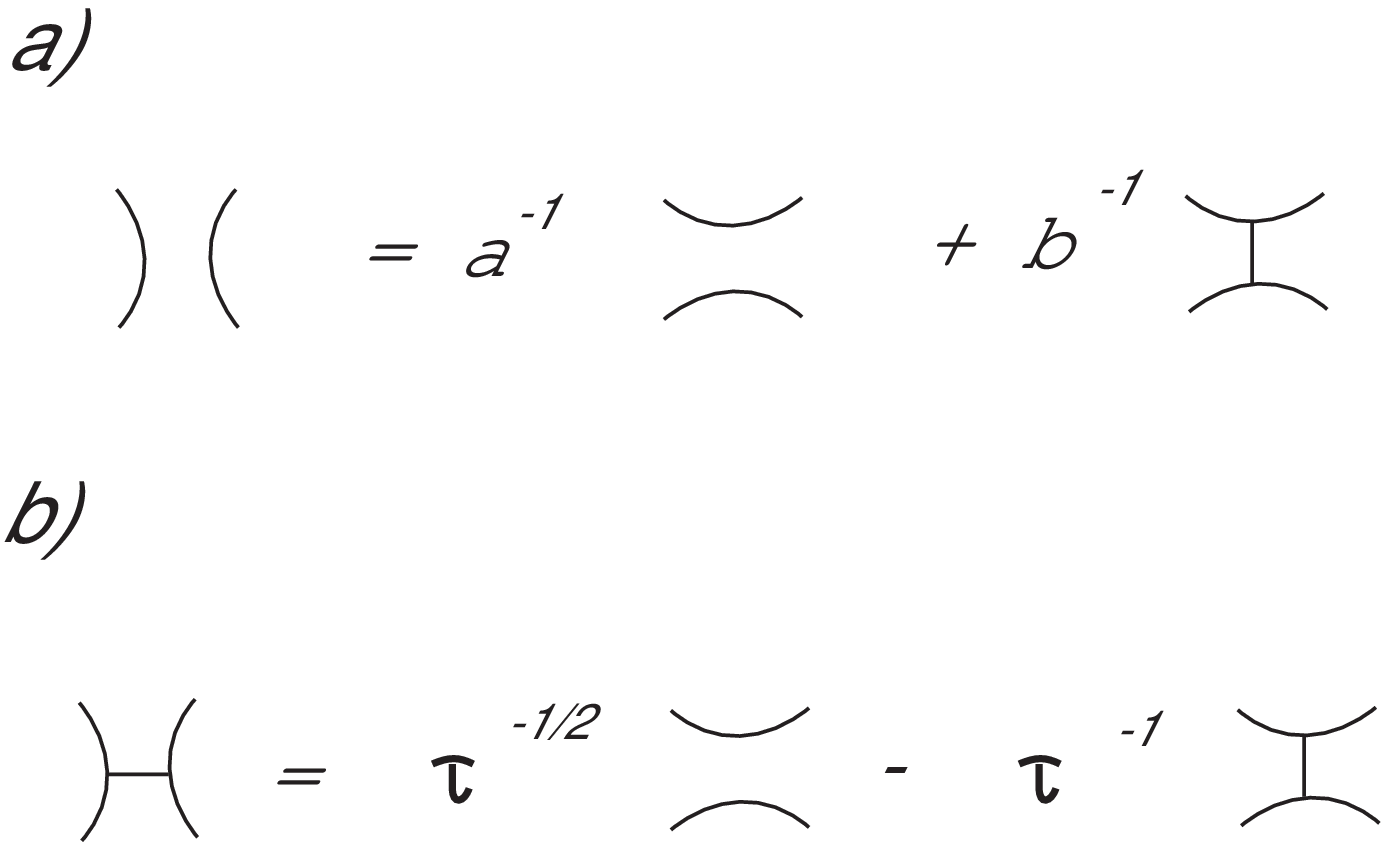}
\caption{The $6j$ symbols can be obtained from the $h=0,i=1$
and $h=1, i=a^{-1}$ relations.}
\label{fig:psnf5}
\end{figure}

In order to calculate the $6j$-symbol, we set $h=0,i=1$ in $R$ to get
the equation in fig.~\ref{fig:psnf5} a).
Unitarity requires ${\mid a^{-1} \mid}^2 + {\mid b^{-1} \mid}^2 = 1$ so
$a=\tau$ and $b=e^{2 \pi i \phi} \sqrt{\tau}$ where $\phi$ is an irrelevant phase
(associated with identifying the simplest 3-point diagram with some unit vector
in $V(D,3)$) which we set to $1$.

Similarly setting $h=1, i=a^{-1}$ (and hence $x=0$) we find the second row of the $6j$-symbol (i.e. the $F$ matrix) in fig.~\ref{fig:psnf5} b) and thus:
\begin{equation}
F = \begin{vmatrix}
\tau^{-1} & \tau^{-1/2} \\
\tau^{-1/2} & -\tau^{-1}
\end{vmatrix}
\end{equation}

Thus, fixing $V(D,4)$ to have minimal dimension, $2$, we find our constants specified:
$a=\tau, b=\tau^{1/2}, c=-\tau^{-1/2}$, and the $F$-matrix as well.

What we have obtained is the Turaev-Viro (or ``doubled") version of the unitary
Fibonacci fusion category.  From these rules -- $a,b,c,$ and $F$ -- all nets $G$ on
a sphere can be evaluated to a scalar ${\langle G \rangle}_\tau$ which is the ``golden''
quantum invariant.  Similarly, with this data the entire unitary modular functor is
specified on all surfaces with or without boundary: $V \cong $ DFib.

Remark: If $V(D,4)$ is allowed to have dimension $3$, a generic solution, the Yamada polynomial, exists.  It can be truncated to doubled $SO(3)_k$ - modular functors for $k$ odd $>3$ by imposing the correct dimension restriction on $V(D,2k)$.  If $V(D,4)$ is allowed to be $4$-dimensional, a relation in $V(D,5)$ realizes Kupperberg's $G_2$-spider which presumably admits further specializing relations which generate $G_2$ level $k$ topological
quantum field theories (TQFTs).

\section{Deriving the Properties of DFib from $V(\Sigma,n)$}

In the remaining three sections of this paper, we will
discuss the properties of the ground states $V(\Sigma,n)\subset {\cal H}(\Sigma,n)$,
thereby obtaining physical properties of the topological phase DFib.
DFib is the the product of two copies of opposite chirality of the
Fibonacci theory, Fib, which the simplest $2+1$ dimensional TQFT
theory with nonabelian braiding rules.
(Fib is also the simplest universal theory \cite{FLW}.)
It arises as the ``even sub-theory''
of $SU(2)_3$ (i.e integer spins only) or $SU(3)_2$ and
also directly from $G(2)_1$.

Fib has one non-trivial particle $\tau$ with fusion rule:
\begin{equation}\tau \otimes \tau = 1 \oplus \tau \end{equation}
Of course the quantum dimension of $\tau$ is $\frac{1 + \sqrt{5}}{2} = \tau$.
(In a slight abuse of notation, we use $\tau$ to denote both the
particle and its quantum dimension, the golden ratio.)

A discrete manifestation of this quantum dimension is that Fib$(S^2, n+2)$, the Hilbert space for $n+2$ $\tau$-particles at fixed position on the $2$-sphere, is Fib$(n)$, the $n$th Fibonacci number.  (Proof: fuse two of the particles: the result will be either $n+1$ or $n$ $\tau$'s on $S^2$ depending on the fusion process outcome.  This yields the famous recursion formula: dim Fib$(S^2,n+2) =$ dim Fib$(S^2,n+1) +$ dim Fib$(S^2,n)$ which defines Fibonacci numbers.)  This gives the exponential growth referred to in the introduction.

Fib is a chiral theory with the following parameters:
\begin{equation} S= \frac{1}{\sqrt{\tau+2}} \begin{vmatrix} 1 & \tau \\ \tau & -1 \end{vmatrix}
\end{equation}
\begin{equation}S^\tau_{\tau \tau} = e^{3 \pi i / 10} \end{equation}
\begin{equation} F=\begin{vmatrix} \tau^{-1} & \tau^{-1/2} \\ \tau^{-1/2} & - \tau^{-1} \end{vmatrix} \end{equation}
\begin{figure}[tbh!]
\includegraphics[width=2in]{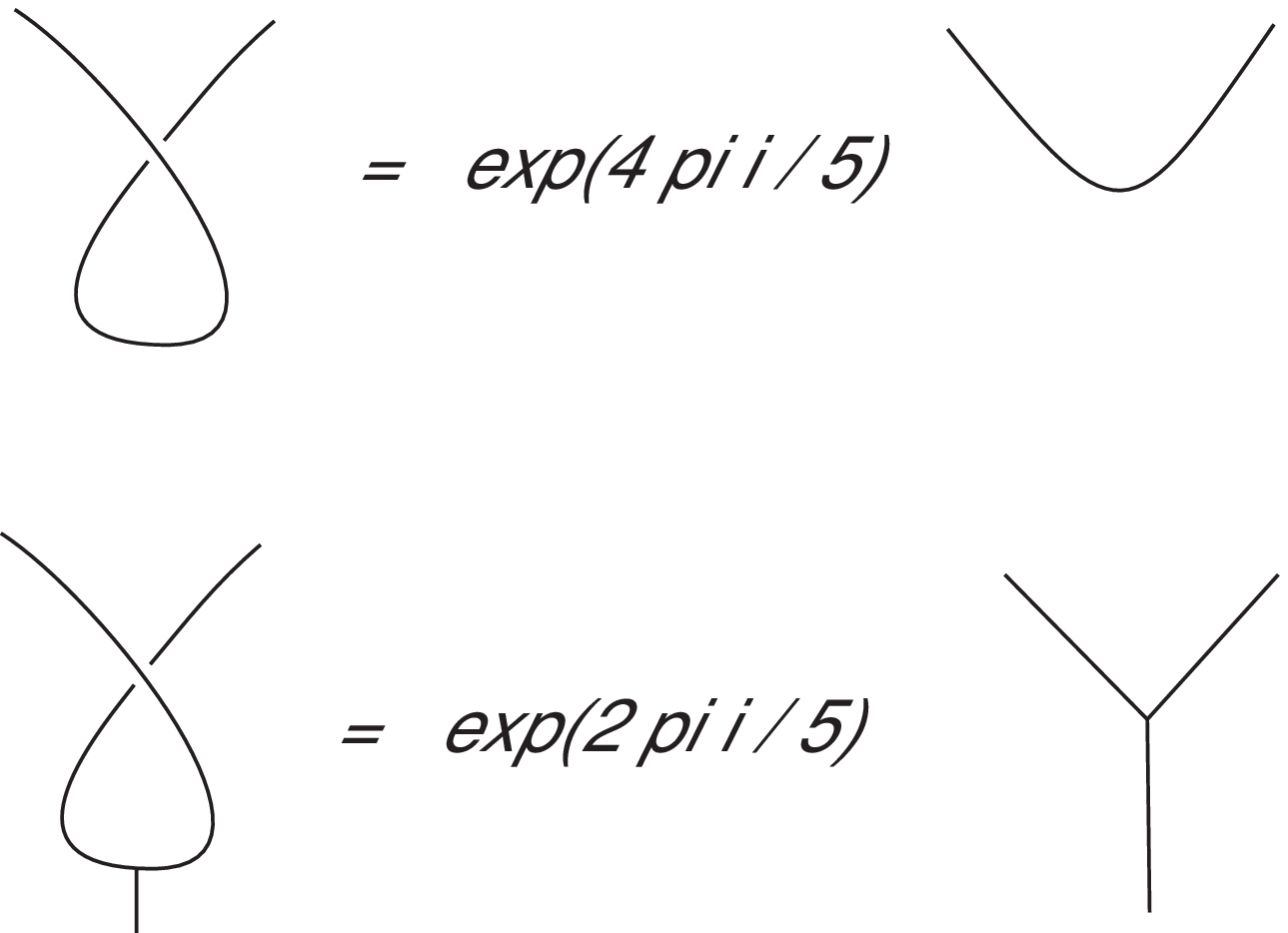}
\label{fig:psnf6}
\end{figure}

The theory $V$ constructed in section $III$ is isomorphic to $\text{Fib}^* \otimes \text{Fib} \cong \text{End} (\text{Fib})$.  Very briefly we explain this connection in the context of a closed surface $\Sigma$ (visualize $\Sigma =$ torus).  Let $G$ be a fine net on $\Sigma$ which ``fills it" in the sense that all complementary regions are disks $\{ \delta_i \}$.  According to Reshitikin-Turaev \cite{RT},
labelings (by $1$ or $\tau$) of the bonds of $G$ consistent with
Fib fusion rules span the Hilbert space for a very high genus
surface $\Sigma^\dag$ which is the boundary of $\bar{G}$,
($\bar{G}$ is a $3$-D thickening of the net $G$).  It can be seen that the fixed space under the $F$-matrix action on $G$-labelings can be obtained from projectors associated to $\{ \delta_i \}$.  These disks determine projectors, or plaquette operators, onto the trivial particle type along a collection of ``longitudes'' on $\Sigma^\dag$ (where $\delta_i$ intersects $\Sigma^\dag$).  Another way to implement these plaquette operators is to add $\{ \bar{\delta_i} \}$ (thickenings of $\{ \delta_i \}$ to $\bar{G}$).  $\bar{G} \bigcup \{ \bar{\delta_i} \}$ is homeomorphic to a product $\Sigma \times I$, surface cross interval.  Adding the plaquet operators has cut Fib$(\Sigma^\dag)$ down to
Fib$(\partial (\Sigma \times I)) =$ Fib$(\bar{\Sigma} \coprod \Sigma) =
\text{Fib}^*(\Sigma) \otimes \text{Fib}(\Sigma) = \text{DFib}(\Sigma)$.

From this point of view the double arises from the fact that $\text{surface} \times I$ has both an ``inner" and ``outer" boundary.  It is nontrivial to align the various structures (e.g. particles) of DFib from the two perspectives: one as a theory of trivalent graphs on a surface (Section III) and the other as a tensor product of a chiral theory and its dual.  For this reason we are not content to merely state that the particle content of DFib is $1 \otimes 1, 1 \otimes \tau, \tau \otimes 1$, and $\tau \otimes \tau$.  Rather, we will give a direct string
representation for these particles shortly.

\begin{figure}[tbh!]
\includegraphics[width=2in]{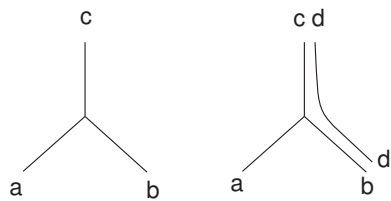}
\caption{The addition of a passive arc to the (piece of a) string net on the right
preserves the rules satisfied by nets.}
\label{fig:psnf7}
\end{figure}

{\it Remark:}  The expression of DFib (and more genreally, Turaev-Viro theories) through commuting local projectors was known to Kitaev and Kuperberg and made explicit for
DFib in ref.~\onlinecite{LWstrnet}.  A conceptual understanding of the commutation relations is readily at hand from the preceding construction of $\Sigma \times I$.  The ``longitudes'' on which we apply plaquette projectors are disjoint and thus commuting.  The fusion rules that are enforced at disjoint vertices also commute.  Finally vertex and plaquet terms commute because rules are preserved under the addition of an additional (``passive'') particle trajectory, labeled $d$ in fig.~\ref{fig:psnf7}.

We conclude this section by finding the $4$ irreducible representations of the DFib annulus category.  This is the linear $C^*$-category whose objects are finite point sets (boundary data) on a circle and whose morphisms are formal combinations of nets in the annulus, which obey the linear rules $a,b,c,$ and $F$, and which mediate between the boundary data.

The four irreducible category (or ``Algebroid'') representations are detected as idempotents in $A_{0,0}$ and $A_{1,1}$, the algebra under stacking of nets in annuli with either trivial or $1$-point boundary data.  A table which organizes the results and compares back to the
$\text{Fib}^* \otimes \text{Fib}$ picture is given in the table below. 
The entries show the dimensions of the Hilbert space of (formal combinations of) nets on an annulus which start on the inner boundary with a given boundary condition (horizontal axis) and terminate near the outer boundary with a copy of a given idempotent (vertical axis).

\vspace{7mm}

\begin{center}
\begin{tabular*}{0.45\textwidth} {@{\extracolsep{\fill}}ccccc}
&\multicolumn{4}{l}
        {\rule[-3mm]{0mm}{8mm} boundary conditions} \\
    irreps (idempotents)   &     $   0$ & $1$          & $2$       & \ldots \\ \hline \hline
$e_1 \cong 1 \otimes 1$    & $1$                   & $0$          & $> 0$    &    \\ 
$e_2 \cong \tau \otimes \tau$ & $1$   & $1$ & $>0$ & $ > 0$\\ 
$e_3 \cong 1 \otimes \tau$ & $0$     & $1$  & $>0$ & \\ 
$e_4 \cong \tau \otimes 1$ & $0$  & $1$ & $>0$ & \\
\label{table1}
\end{tabular*}
\end{center}

\vspace{7mm}

Let us start by finding the idempotents for $A_{0,0}$.  $A_{0,0}$ has the empty net as its identity and is generated by a single ring $R$.
\begin{figure}[tbh!]
\includegraphics[width=2in]{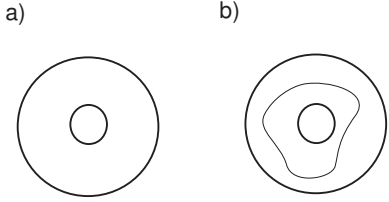}
\caption{a) $1 \in A_{0,0}$.  b) $R \in A_{0,0}$.}
\label{fig:psnf9}
\end{figure}
The $a,b,c,F$ rules show: $R^2 =1+R$ so $A_{0,0} \cong C[R] / (R^2 = 1+R)$.

\begin{figure}[tbh!]
\includegraphics[width=2.5in]{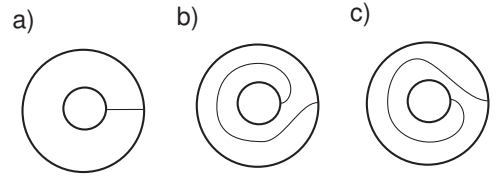}
\caption{a) $1 \in A_{1,1}$, b) $T \in A_{1,1}$, c) $T^{-1} \in A_{1,1}$}
\label{fig:psnf10}
\end{figure}

\begin{figure}[tbh!]
\includegraphics[width=1.7in]{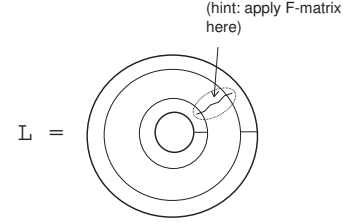}
\caption{A compact representation of $L = \tau^{-1/2} 1 + \tau^{-3/2} (T + T^{-1})$,
as may be seen by applying the $F$-matrix where indicated.}
\label{fig:psnf11}
\end{figure}

\begin{figure}[tbh!]
\includegraphics[width=2in]{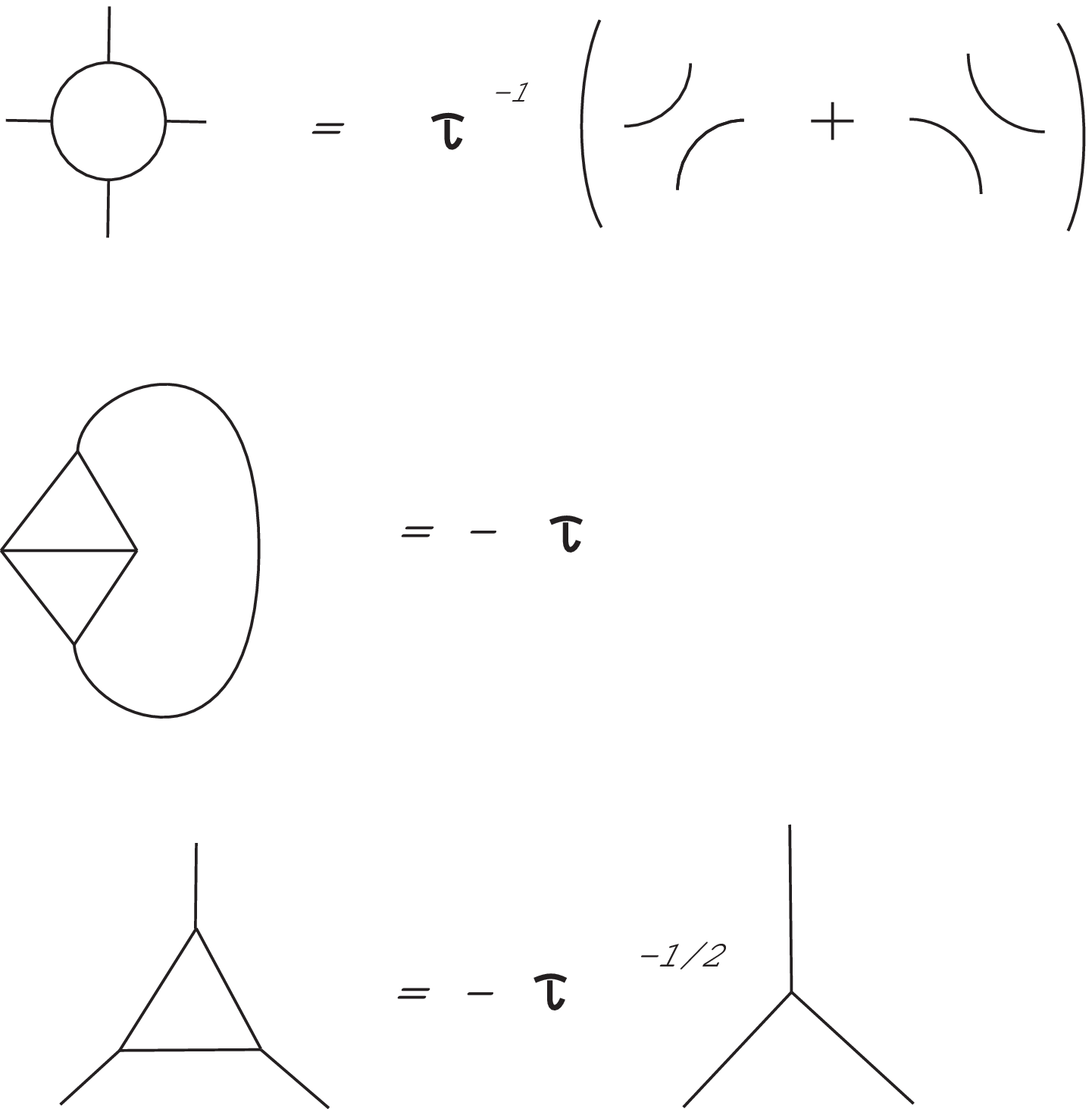}
\caption{The relations depicted above can be obtained by applying the $a,b,c,$ and $F$ rules.}
\label{fig:psnf12}
\end{figure}

More generally, the idempotents in the algebra $C[x] / P(x)$, $P(x)=(x-a_1) \ldots (x-a_k)$, all roots distinct, are given by:
\begin{equation}
\label{eqn:idem}
e_i = \frac{(x-a_i) \ldots \widehat{(x-a_i)} \ldots (x-a_k)}{(a_i-a_1) \ldots \widehat{(a_i-a_i)} \ldots (a_i - a_k)}
\end{equation}
We get
\begin{eqnarray}
e_1 &=& \frac{1+\tau R}{\tau+2} \text{      (} \cong 1 \otimes 1, \text{the trivial particle} \text{)} \\
e_2 &=& \frac{1+\bar{\tau}R}{2+\bar{\tau}} \text{       (} \cong \tau \otimes \tau \text{).}
\end{eqnarray}
$\bar{\tau}=-\tau^{-1}$.  Using the $a,b,c$, and $F$ rules the algebra $A_{1,1}$ is seen to be generated by the identity $1$, $T$, and $T^{-1}$ (where $T$ is defined in Figure \ref{fig:psnf10}).
There is an element $L = \tau^{-1/2} 1 + \tau^{-3/2} (T + T^{-1})$ which factors in a category sense through $A_{0,0}$.  In fact, using the the elementary rules in Figure \ref{fig:psnf3} we derive the useful identities in Figure \ref{fig:psnf12}
and we find that $L$ is equivalently represented as shown in Figure \ref{fig:psnf11}.
It follows that to find a new representation of the annulus category we should
look for the idempotents in $A_{1,1}/L$.  $A_{1,1}/L \cong C [T] / \{T^2 + \tau T + 1 = 0 \}$. 
So, using (\ref{eqn:idem}) again and a little manipulation we find idempotents
\begin{equation}
\tilde{e}_3 = \frac{T-\left( \frac{\tau - \sqrt{\tau-3}}{2} \right) 1}{\left(-\sqrt{\tau-3} \right)}
\end{equation}
and
\begin{equation}
\tilde{e}_4=\frac{T+\left( \frac{\tau+\sqrt{\tau-3}}{2} \right) 1}{\sqrt{\tau-3}}
\end{equation}
in the quotient algebra.  

Also $L^2 = (\tau^{1/2} + \tau{-3/2}) L$ so $e_L = L / {\tau^{1/2} + \tau^{-3/2}}$
so again using the identities in Figure \ref{fig:psnf12} we can find $e_3 = \tilde{e}_3 (1-e_L)$
and $e_4 = \tilde{e}_4 (1-e_L)$:
\begin{equation}
e_3 = \tilde{e}_3 + \frac{\tau+2+\sqrt{\tau-3}}{2(\tau^{1/2} + \tau^{-3/2}) \sqrt{\tau-3}} L
\end{equation}
\begin{equation}
e_4 = \tilde{e}_4 + \frac{-(\tau+2) - \sqrt{\tau-3}}{2(\tau^{1/2} + \tau^{-3/2}) \sqrt{\tau-3}}L
\end{equation}

\section{Chromatic Polynomial and Yamada Polynomial}

\begin{figure}[tbh!]
\includegraphics[width=2.5in]{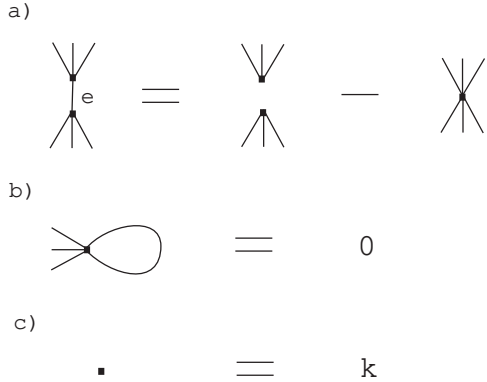}
\caption{(a) Graphical depiction of the ``delete-contract'' recursion relation.
This relation together with the two depicted graphically in (b) and (c) and
multiplicativity under disjoint union completely
determine $\chi_{\hat{G}}(k)$.}
\label{fig:psnf13}
\end{figure}

The chromatic polynomial $\chi_{\hat{G}}(k)$ of a graph $\hat{G}$ at the positive integer $k$ counts the number of $k$-colorings of the vertices of the graph (so that no two vertices connected by a bond are given the same color).  $\chi$ obeys the famous ``delete-contract'' recursion relation:
\begin{equation}
\chi_{\hat{G}} (k) = \chi_{\hat{G} - e} (k) - \chi_{\hat{G}/e}(k)
\label{eqn:delcon}
\end{equation}
This relation can be depicted graphically as shown in figure (\ref{fig:psnf13}),
in which we have suppressed $\chi$ (as we have consistently suppressed
the wave function and written the relation out in terms of its pictorial argument).
This is a local relation; there is no significance to the bit of $\hat{G}$
near the active edge $e$, which in the drawing is represented,
purely for illustrative purposes, by three half-edges on top and bottom.
$\chi(k)$ is completely fixed by (\ref{eqn:delcon}) and the following conditions:
that $\chi$ vanish on any graph in which the two ends of a single bond are joined
to the same vertex, that $\chi_{\text{single pt.}} (k) = k$, and multiplicativity under
disjoint union:
\begin{equation}
\chi_{\hat{G} \coprod \hat{H}} = \chi_{\hat{G}} \chi_{\hat{H}}
\end{equation}
We will be interested in $\chi_{\hat{G}}$ evaluated at noninteger values as well as integral ones.

\begin{figure}[tbh]
\includegraphics[width=2.5in]{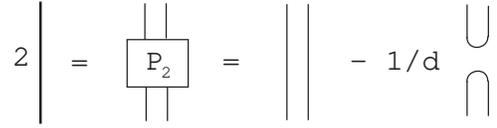}
\caption{Jones-Wenzl projector.}
\label{fig:psnf14}
\end{figure}
We now turn to the Yamada polynomial defined for a net (trivalent graph) $G$
lying in the plane (or 2-sphere); we denote it by
${\langle \langle G \rangle \rangle}_d$,
where $d$ is the variable.  To define ${\langle \langle G \rangle \rangle}_d$,
recall the 2-strand Jones-Wenzl projector,
an idempotent familiar from the study of the Temperley-Lieb algebra $TL_d$.
Given $G$, ${\langle \langle G \rangle \rangle}_d$ is defined by labeling every arc of $G$ by $2$ as in (\ref{fig:psnf14}) and then expanding to a weighted superposition of multi-loops.  In each term, each loop contributes a numerical factor of $d$,
\begin{equation}
{\langle \langle G \rangle \rangle}_d = \sum_\text{terms} (\text{coeff.}) d^{\text{\# loops}}.
\end{equation}

\begin{figure}[tbh!]
\includegraphics[width=3in]{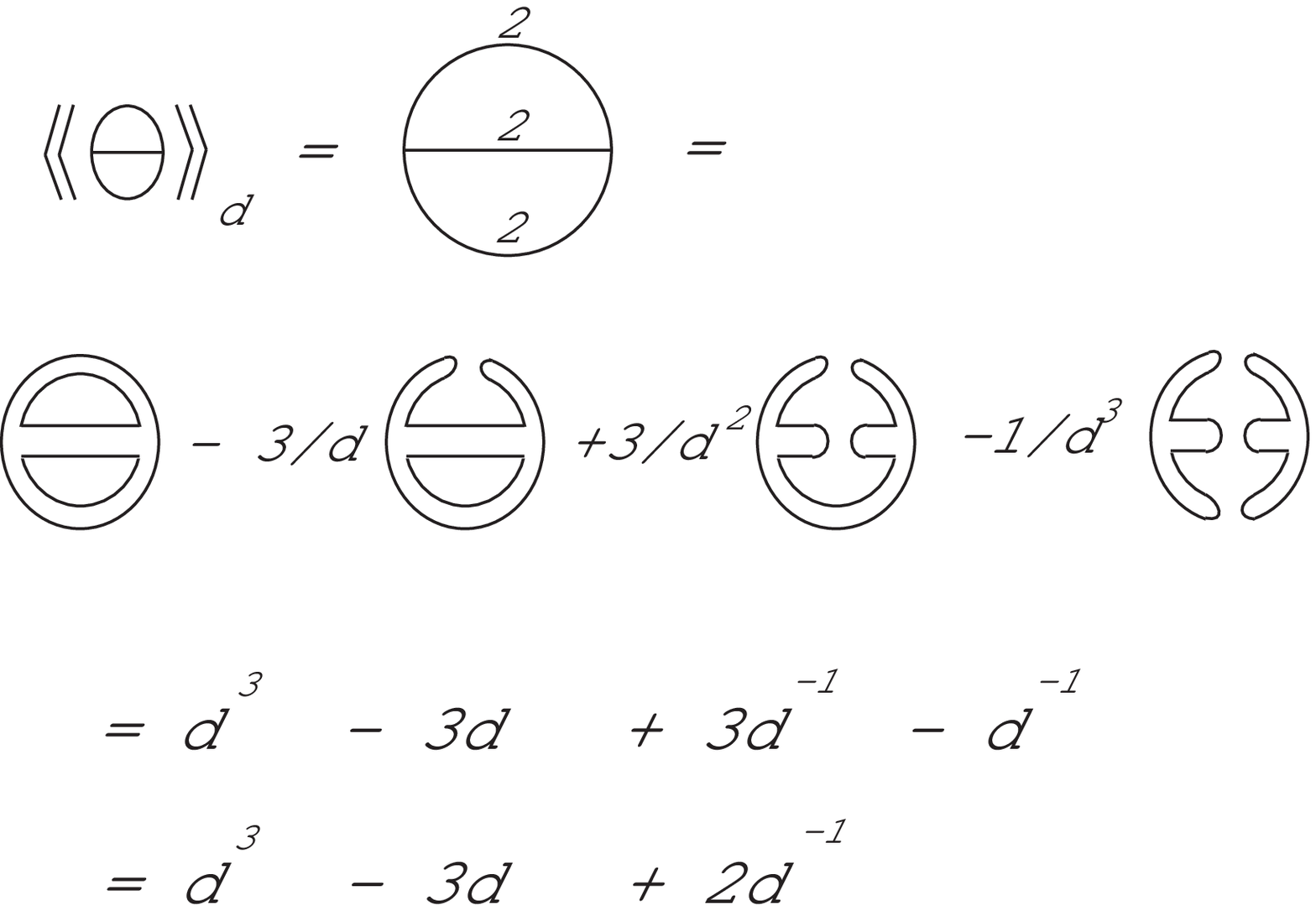}
\caption{The Yamada polynomial for the $\theta$ graph.}
\label{fig:psnf15}
\end{figure}

For example, if $G$ is a graph shaped like the Greek letter $\theta$,
we have the result shown in Figure \ref{fig:psnf15}. 
As this example shows, the Yamada polynomial is actually a
polynomial in $d$, $d^{-1}$.

\newtheorem{thm2}{Theorem}
\begin{thm2}
If $G$ is a net in the $2$-sphere and $\hat{G}$ is the dual graph then:
\begin{equation} {\langle \langle G \rangle \rangle}_d = d^{-V(\hat{G})} \chi_{\hat{G}} (d^2) \end{equation}
$V = $ number of vertices of $\hat{G}$ = number of faces of $G$.
\end{thm2}

\begin{figure}[tbh!]
\includegraphics[width=2.5in]{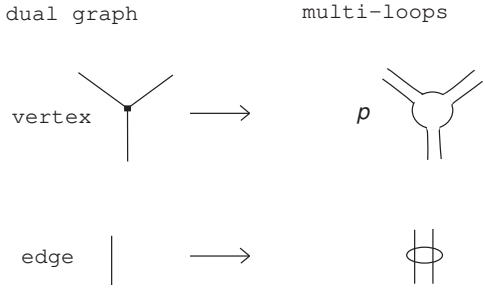}
\caption{Local rules for turning the dual graph of a net into a superposition of multi-loops.  The ellipse represents an unknown combination}
\label{fig:psnf15a}
\end{figure}

\begin{figure}[tbh!]
\includegraphics[width=2in]{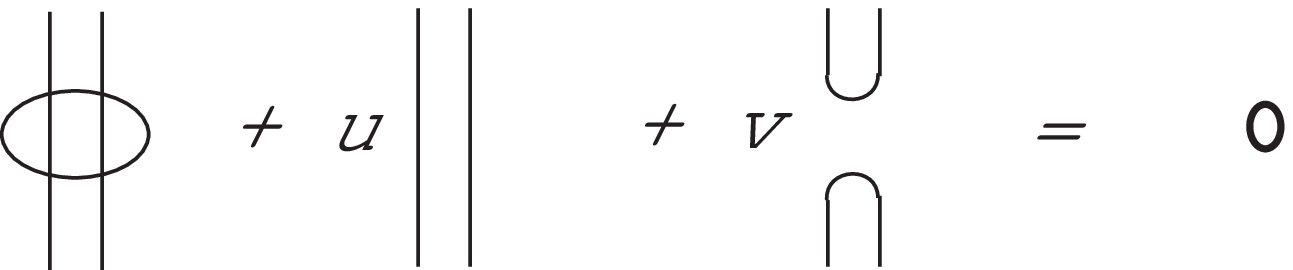}
\caption{A recoupling rule by which the dual of a trivalent graph can be converted into multi-loops.
We will choose $u$ and $v$ (and also $\rho$) so that the procedure for
making a trivalent graph into a multi-loop relates $\chi_{\hat{G}} (k)$ and
${\langle \langle G \rangle \rangle}_d$}
\label{fig:psnf16}
\end{figure}

{\it Proof:}  The above procedure for turning a net into a superposition of multi-loops may be generalized by declaring two local rules shown in fig.~\ref{fig:psnf15a}.
The first rule says a point is replaced by a circle with a possible numerical weight (vertex fugacity) and each arc is replaced by two lines with some general recoupling.  We express the fact that this recoupling is still a variable to be solved for by Figure \ref{fig:psnf16}.
Our goal is to find suitable values for $\rho$, $u$, and $v$ so that $\chi_{\hat{G}} (k)$ comes out related to ${\langle \langle G \rangle \rangle}_d$.  The correct $k$ will turn out to be $d^2$.

\begin{figure}[tbh!]
\includegraphics[width=2.5in]{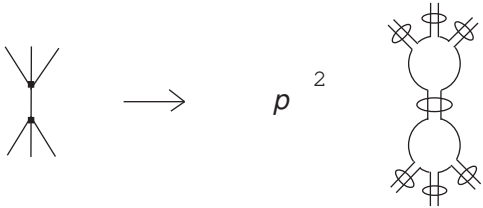}
\caption{Applying the rules in fig.~\ref{fig:psnf15a} to the graph on the left-hand side
of the figure above yields the picture on the right.}
\label{fig:psnf16b}
\end{figure}

Let us look near a typical edge $e$ of $\hat{G}$ and expand it using the rule in
fig.~\ref{fig:psnf16b}.

\begin{figure}[tbh!]
\includegraphics[width=3in]{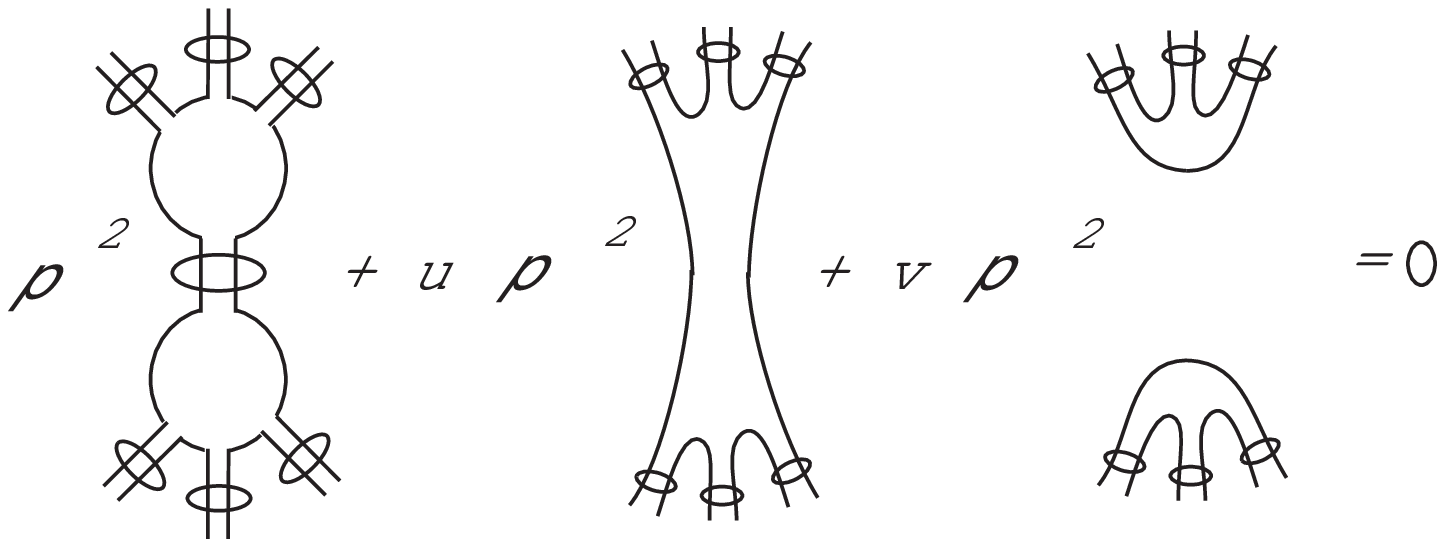}
\caption{Using the recoupling rule in Figure \ref{fig:psnf16}, we can simplify the picture on the right-hand-side
of fig.~\ref{fig:psnf16b}.}
\label{fig:psnf16c}
\end{figure}

\begin{figure}[tbh!]
\includegraphics[width=3in]{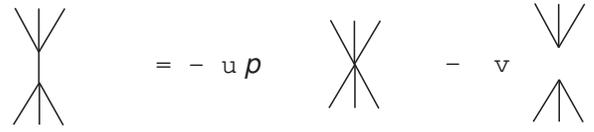}
\caption{A relation that must be satisfied by $u$, $v$, and $\rho$.}
\label{fig:psnf16d}
\end{figure}

\begin{figure}[tbh!]
\includegraphics[width=3.25in]{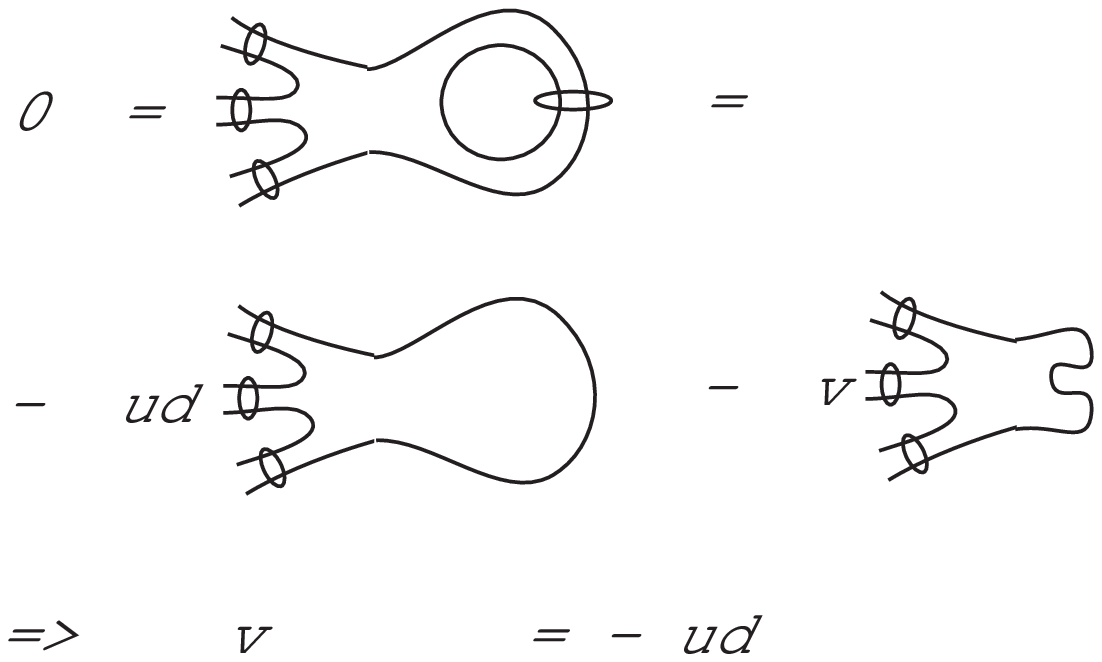}
\caption{A graph with an edge that connects a vertex to itself evaluates to zero,
from which we deduce a relation between $u$ and $v$.}
\label{fig:psnf17}
\end{figure}

With this expansion and fig.~\ref{fig:psnf16} we obtain fig.~\ref{fig:psnf16c}.
 Translating this back into graphs yields fig.~\ref{fig:psnf16d}.
 Also, since graphs with an edge that connects a vertex to itself evaluate to zero,
we have fig.~\ref{fig:psnf17}, implying $v=-ud$.  Also, $k = \rho d$.

\begin{figure}[tbh!]
\includegraphics[width=3in]{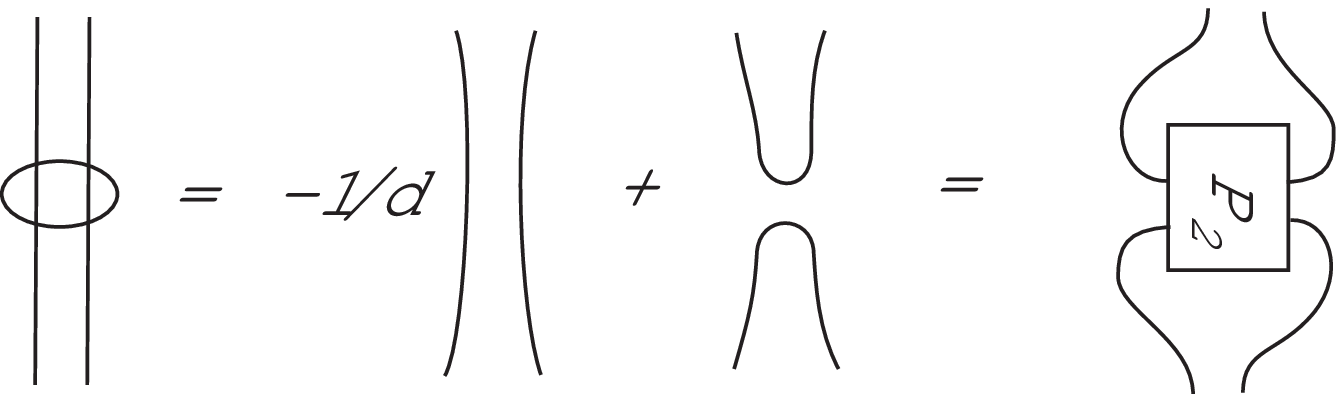}
\caption{Solving for $u,v,\rho$, we see that the recoupling rule
is just a ``sideways'' $P_2$.}
\label{fig:psnf17b}
\end{figure}

Comparing fig.~\ref{fig:psnf16d} with the chromatic relation (\ref{eqn:delcon}) 
we find: $v=-1, u=1/\rho$.  Because we have $v = -ud$, we obtain $u=1/d$ so $d=\rho$. 
Finally, since $k = \rho d$, $k=d^2$.  The mystery combination turns out to be
a ``sideways'' $P_2$  as shown in fig.~\ref{fig:psnf17b}.

\begin{figure}[tbh!]
\includegraphics[width=2.5in]{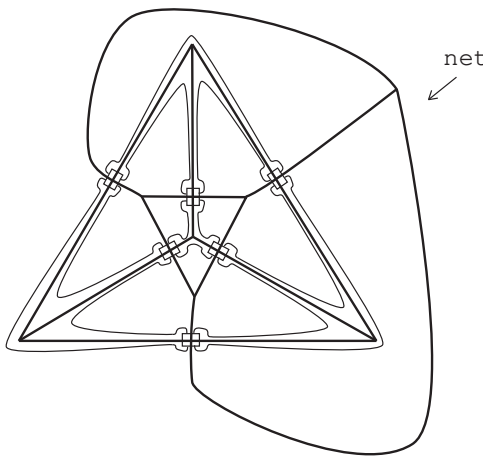}
\caption{Applying the rules to a complete graph.}
\label{fig:psnf18}
\end{figure}

To complete the proof, it remains to see the global geometry of how these
$P_2$'s hook together.  We claim that they lie along (doubled) dual graph edges.
It suffices to examine an example.  We take $\hat{G}$ and $G$ to be the
complete graph $C_4$ shown in fig.~\ref{fig:psnf18}.
The factor of $\rho^{V(\hat{G})}$ appearing with
${\langle \langle G \rangle \rangle}_d$ has been put on the
right-hand-side in the theorem statement.

It is well known that the Yamada polynomial and the unitary invariant ${\langle G \rangle}_\tau$ of Section III are closely related.  Specifically, when $d=\tau$, we should modify the Yamada polynomial
${\langle \langle G \rangle \rangle}_\tau$ by a vertex fugacity to obtain ${\langle G \rangle}_\tau$.  This can be seen by checking that both quantities satisfy the elementary rules of Section III that determine them uniquely.  From our example,
shown in \ref{fig:psnf15}, we find ${\langle \langle \theta \rangle \rangle}_\tau = \tau^{-1}$
while for the unitary theory ${\langle \theta \rangle}_\tau = ab = \tau^{3/2}$.
To convert from the unnormalized (Kauffman) theory to the unitary theory
(compare to ref.~\onlinecite{LWstrnet}) one must multiply in a factor of $\tau^{5/4}$ for each vertex.
Thus:
\begin{equation}
\label{eqn:norm}
{\langle G \rangle}_\tau = (\tau^{5/4})^{V(G)} \tau^{-V(\hat{G})} \chi_{\hat{G}} (\tau^2)
\end{equation}
Using the Euler relation $F(\hat{G}) + V(\hat{G}) - E(\hat{G})$ and the fact
that $\hat{G}$ is a triangulation, implying $E=\frac{3}{2} F$, we find:
\begin{equation}
F(\hat{G}) = 2V(\hat{G}) - 4
\end{equation}
Therefore:
\begin{equation}
{\langle G \rangle}_\tau = \tau^{-5} \tau^{\frac{3}{2}
V(\hat{G})} \chi_{\hat{G}} (\tau + 1) \label{eqn:tp1}
\end{equation}
We can now use Tutte's ``golden ratio theorem" ([T] and [L]):
For a planar triangulation $\hat{G}$:
\begin{equation}
\label{eqn:tutte1}
(\chi_{\hat{G}} (\tau+1))^2 \tau^{3V} (\tau+2) (\tau^{-10}) = \chi_{\hat{G}} (\tau+2)
\end{equation}
This formula allowed Tutte to conclude that the r.h.s. is positive,
creating a curious analogue (and precursor) to the $4$-color theorem.
(Neither result has been shown to imply the other.)

Now square (\ref{eqn:tp1}) and substitute into (\ref{eqn:tutte1});
the result is a remarkable formula:

\begin{equation}
\label{eqn:tutte2}
({\langle G \rangle}_\tau)^2 = \frac{1}{\tau+2} \chi_{\hat{G}} (\tau+2)
\end{equation}

We have just proved the formula when $\hat{G}$ is a triangulation;
in fact it holds more generally whenever $G$ is a net. 
To establish this it suffices to check the formula when $G$ is a single loop
and to observe that the formula behaves well under disjoint union of
disconnected components of $G$: the l.h.s. is obviously multiplicative and,
it turns out, so is the r.h.s.  The reason is that disjoint union of $G_1$ and $G_2$
corresponds to a 1-point union $\hat{G}_1 \bigvee \hat{G}_2$, and with the
factor of $1/(\tau+2)$, the chromatic polynomial becomes multiplicative under $1$-point unions.
So we have proved:

\newtheorem{thm3}{Theorem}
\label{thm:thm3}
\begin{thm3}
Let $G$ be a net in the plane or $2$-sphere (possibly disconnected and possibly with circle components).  Then
\begin{equation}
\label{eqn:tutte3}
{{\langle G \rangle}_\tau}^2 = \left( \frac{1}{\tau+2} \right) \chi_{\hat{G}}(\tau+2)
\end{equation}
\end{thm3}
{\it Note:}  We would like to thank P. Fendley and E. Fradkin for a
discussion of this identity.  In their paper \cite{FF}, a non-unitary
normalization for  ${\langle G \rangle}_\tau$ led to
a vertex fugacity on the right-hand-side of (\ref{eqn:tutte3}),
thereby obscuring the simplicity of (\ref{eqn:tutte3}) and the direct connection to the
Potts model.

\section{String Net Wavefunctions and the Potts model}

Let us review the high- and low-temperature expansions of the Potts model. 
We begin by assuming a lattice with $Q$ ``spin values'' ${\sigma_i}=0,1,\ldots,Q-1$
at each site $i$.
The partition function $Z(Q)$ is defined by:
\begin{eqnarray}
Z &=& \sum_{\sigma}
\exp\left(-\beta \left( -J \sum_{\langle i,j \rangle}
\delta_{\sigma_i, \sigma_j} \right) \right) \\ &=&
\sum_{\sigma} \prod_{\langle i, j \rangle}
\exp \left( \beta J \delta_{\sigma_i, \sigma_j} \right)
\end{eqnarray}
where the sum is over spin state configurations.
Setting $\gamma = e^{\beta J} - 1$ and expanding in powers of $\gamma$ we obtain
the high-temperature expansion:
\begin{equation}
\label{eqn:hte} Z=\sum Q^{c}
\:\: \gamma^{b}
\end{equation}
where the sum is over bond configurations;
$c$ is the number of clusters, and $b$ is the number of bonds.
From now on, we consider the ferromagnetic case $J=1$.
Note that the last sum is over the $2^b$ distinct subsets,
not the $Q^{\text{(\# of sites)}}$ distinct colorings because
a Fubini resummation has taken place. 
Also, note that isolated sites count as clusters in (\ref{eqn:hte}).
This is the Fortuin-Kateleyn representation\cite{FK}.
It is known that for $0 < Q \leq 4$ the model is critical precisely
at its self-dual point, $\gamma = \sqrt{Q}$.

The high temperature expansion has been used\cite{FF,FNS}
to study loop gases with ground state wavefunction whose amplitude
is $d^{L}$ for some real number $d$, where $L$ is the number
of loops.
The square of such a wave function can be interpreted as a Gibbs
weight $(d^2)^L$ providing a ``plasma analogy''
between topological ground states and the statistical physics of loop gases.
We can easily see that a loop gas with
Gibbs weight $d^2$ per loop is critical if $d \leq \sqrt{2}$:
\begin{equation}
\label{eqn:lg}
(d^2)^L = (d^2)^{c+c^*} = (d^2)^{2c+b}=(d^4)^c (d^2)^b
\end{equation}
Here, $c^*$ is the number of dual clusters, i.e. the minimum number
of occupied bonds which have to be cut in order to make each cluster tree-like
(essentially the number of ``voids'' which are completely contained within clusters).
The first equality follows from each loop being either the outer boundary
of a cluster or a dual cluster.
The second is due to the Euler relation ${c^*} = c+b + \text{const}$.
To map this squared wavefunction to the Potts model,
we need $d^4 = Q \leq 4$; note the edge fugacity for this loop model
is automatically $\sqrt{Q}$, placing the model at its self dual point.

The low-temperature expansion of the Potts model is:
\begin{equation}
Z(Q) = \sum_G \chi_{\hat{G}} (Q) (e^{-\beta J})^L
\end{equation}
where the sum is over graphs $G$ and $L$ is the total length
of the graph. We specialize to the case in which the dual lattice
is trivalent (e.g. the triangular lattice, whose dual lattice is the honeycomb),
so that the graphs will are all trivalent.
At the critical point ($Q \leq 4$), this becomes
\begin{equation}
Z(Q) = \sum_G \chi_{\hat{G}} (Q) \left( \frac{1}{\sqrt{Q} + 1} \right)^L
\end{equation}
using the condition for criticality $e^{{\beta_c} J}-1=\sqrt{Q}$.

The (unnormalized) isotopy invariant wave function $\Psi$ constructed in
Section III satisfies $\Psi(G) = {\langle G \rangle}_\tau$.
Hence the corresponding statistical physics (of equal time correlators)
is that of the normalized probability distribution:
\begin{equation}
\text{prob}(G) = \frac{1}{Z(Q=\tau+2,\beta=0)}\: \left({\langle G \rangle}_\tau\right)^2
\end{equation}
Note that (using (\ref{eqn:tutte3})) this is the high temperature limit of the low temperature expansion.

This situation reminds us of the Toric code \cite{Kitaev97}, where the weight on
loops may be obtained by setting $Q=2$, $\beta=0$ (for $Q=2$, branched nets have zero weight),
so the toric code ground state wavefunction can also be understood
in terms of the high-temperature limit of the low-temperature expansion of
the Potts model. In both cases, we conjecture that
for $\beta < \beta_c$ the wavefunction with isotopy invariance modified by:
\begin{equation}
\Psi(G) = {\langle G \rangle}_\tau x^{\text{length}(G)}
\end{equation}
for $1 \geq x > 1/(\sqrt{Q}+1)$
($\approx 0.345$ when $Q=\tau+2$, and $\approx 0.466$ when $Q=2$)
will be the ground state for some gapped Hamiltonian in the
corresponding topological phase DFib (or Toric code).

Numerical work \cite{Trebst06} already supports this conjecture in the Toric code case. 
Also, note that at $\beta>\beta_c$ (so $x<1/(\sqrt{Q}+1)$) an effective string tension prevents the nets from fluctuating. Thus, the system leaves the topological phase and enters
a ``geodesic phase'' in which small nets dominate.
Recoupling is now unlikely, so the necessary topological relations
are not well enforced and geodesic continuation allows states on the torus to be fairly well guessed
by measurement within a subdisk Ð violating the disk axiom.
Finally, (\ref{eqn:tutte3}) identifies the ground state of the gapped
Levin-Wen \cite{LWstrnet} Hamiltonian: $\Psi(G) = {\langle G \rangle}_\tau$ as
the $\beta=0$ end point of the conjectured family.

In ref.~\onlinecite{FF}, it is conjectured that the unnormalized probability distribution on
nets $({\langle\langle G \rangle \rangle}_d)^2$ should somehow yield the same
statistical physics as the Potts model at an ``effective" $Q_{\text{eff}}$ satisfying:
\begin{equation}
(Q_{\text{eff}} - 1) = (d^2 - 1)^2.
\end{equation}
If we set $d=\tau$ and adjust the $G$-vertex fugacity so as to replace
${\langle \langle G \rangle \rangle}_\tau$ with ${\langle G \rangle}_\tau$,
as we did in Section V, then the conjecture is precisely verified at $Q=\tau+2, \beta=0$:
\begin{equation}
(\tau+2-1) = (\tau^2 - 1)^2
\end{equation}

We close this section with an observation whose importance, if any,
we do not yet understand.  The high temperature expansion weights loops
while the low temperature expansion weights nets.  But loops are nets,
so we might ask: at what value of $Q$ do the high and low temperature
expansions weigh loops equally?  In the low temperature expansion,
loops have weights $Q-1$ and in the high temperature expansion
they have weight $\sqrt{Q}$ (from \ref{eqn:lg}).  So loops are equally weighted when
$Q-1=\sqrt{Q}$, i.e. when $Q=\tau^2$.

\section{Conclusions}

DFib is among the simplest conceivable achiral particle theories. 
In some sense it rivals the toric code in simplicity (both have $4$ particles),
but is vastly richer (in fact, universal\cite{FLW}) in its braiding. 
We have explored the path to this phase, taking a Hilbert space
${\cal H}$ of nets and an isotopy invariant Hamiltonian
(possibly with a bond fugacity, see Section VI) 
$H: {\cal H} \rightarrow {\cal H}$ as our starting point.
The exploration has been combinatorial in Section III, algebraic in Section IV,
and statistical in Section VI.

In Section III, we saw that DFib emerges from minimizing certain degeneracies.
This encourages us to believe the phase will ultimately be found in
nature - as nature abhors a degeneracy.

In Section VI, we establish that the nets $G$ with (squared) topological weighting
$({\langle G \rangle}_\tau)^2$ (as usual, squaring the wavefunction
to obtain a probability)
are in a high temperature phase of the $(\tau+2)$-state
Potts model (above criticality).  This is also encouraging:
the classical critical point looks as if it is the ``plasma analogy" of
a quantum critical point sitting at the entrance to the DFib phase.
A parallel is explored between this situation and the $Q=2$
Potts critical point which serves as an entrance to the toric code phase.

Unresolved is what, more precisely, is required of
$H: {\cal H} \rightarrow {\cal H}$ to be in the DFib phase.
Is enforcement of the net $G$ structure (encoded in the definition of ${\cal H}$)
plus strong dynamic fluctuation of $G$ adequate?  We do not know.

\begin{acknowledgments}
We would like to thank Paul Fendley and Eduardo Fradkin for useful discussions.  C.N. would like to acknowledge the support of the NSF under grant no. DMR-0411800 and the ARO under grant W911NF-04-1-0236 (C.N.).  This research has been supported by the NSF under grants DMR-0130388 and DMR-0354772 (Z.W.).  L.F. would like to acknowledge the support of the NSF under grant no. PHY -0244728.
\end{acknowledgments}

\appendix

\section{The Chromatic Polynomial and Hard Hexagons}

\begin{figure}[tbh]
\includegraphics[width=2.5in]{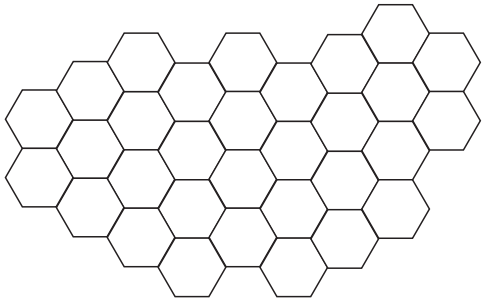}
\caption{large chunk of hexagonal lattice}
\label{fig:psnfaa}
\end{figure}

In this appendix we extend a theorem of Tutte \cite{Tutte} which bounds the decay of the chromatic
polynomial at $\tau+2$ of planar graphs.  Specifically, we obtain a better sharp bound for graphs that
consist of a large regions of hexagonal lattice (see Figure \ref{fig:psnfaa}).
The ground state of the doubled Fibonacci theory DFib consists of a certain superposition of string net configurations on a hexagonal lattice.  As in ref.~\onlinecite{FF} we may construct from it a classical
statistical mechanical model of nets, with the Boltzmann weight of each net equal to the norm squared
of its ground state amplitude.  As discussed in Section V, this statistical mechanical model is the
(solvable) infinite-temperature limit of the $Q=\tau+2$ Potts model.  We are interested in the rough
quantitative behavior of the amplitudes for different types of graphs.  In particular, we can consider a
large chunk of hexagonal lattice, which is just the string net consisting of a large finite region whose bulk
includes every available bond (see Figure \ref{fig:psnfaa}).  We can ask about how its amplitude scales
with the size of the region.  Now, the Boltzmann weight  of a string net is (up to overall scaling) just the
chromatic polynomial at $\tau+2$ of the graph dual to the net, which can be interpreted as the
zero=temperature limit of the $\tau+2$ antiferromagnetic Potts model
defined on the graph dual to the net.
So for the large chunk configuration, we are just interested in the behavior of the zero-temperature free
energy of the $\tau+2$ antiferromagnetic Potts model on the hexagonal lattice.

This model is critical \cite{Saleur} and is expected to be described by a conformal field theory.  On general grounds \cite{Cardy} we expect the free energy to scale like
\begin{equation}
\label{eqn:fe} F = c_0 A + c_1 L + c_2 \text{ log }
L \end{equation}
where $c_2$ is universal and related to the central charge of the CFT and
$c_0$ and $c_1$ are non-universal
(here $A$ is the area, i.e. number of hexagons in the region, and $L$ is the length of the boundary).
It turns out that the model is exactly solvable and we can obtain an exact analytic expression for $c_0$.

To do so, we first use the so-called shadow method \cite{Kauffman} to evaluate $G_\tau$ of a net.  This method works as follows.  We take the net to be located on the sphere and assume that the regions (or faces) that it bounds are simply connected and do not border themselves.  Then the shadow method (applied to DFib) gives $G_\tau$ as a sum over black and white colorings of the faces
\begin{equation} G_\tau = \sum_{\text{colorings C}} F_C {E_C}^{-1} V_C. \end{equation}
The colors black and white are identified with the two particle types of the theory, and
\begin{equation} F_C = \prod_{\text{faces F}} d_{C(F)} \end{equation}
\begin{equation} E_C = \prod_{\text{edges E}} \Theta(C,E) \end{equation}
\begin{equation} V_C = \prod_{\text{vertices V}} \text{Tetrahedron} (C,V). \end{equation}
Here $d_{C(F)}$ stands for the quantum dimension associated with the color of face $F$ ($1$ for black, $\tau$ for white).  $\Theta (C,E)$ is the theta graph along whose three edges run particle types associated with the edge $E$ (always the nontrivial particle) and the two faces which $E$ borders.  Tetrahedron $(C,V)$ is the tetrahedral graph with its six edges labeled by the particle types corresponding to the faces and edges adjoining $V$.

Let $v=\# \text{ vertices}, f=\# \text{ faces}, e=\# \text{ edges}$.  When applied to the net which consists of a large chunk of hexagonal lattice, the shadow method reduces to the following: the region containing the point at infinity is black, and the sum over colorings becomes a sum over colorings in which there are no adjacent black hexagons (and no black hexagons adjacent to the black outside region).  Let us first compute the weight of the all white coloring.  We ignore boundary effects.  We have first of all $v = 2f$ and $e = 3f$.  Each edge contributes $\Theta^{-1} = \tau^{-3/2}$, each vertex Tetrahedron $= -\tau$, and each face the quantum dimension $\tau$.  The total weight of the all white coloring is thus
\begin{equation} \left( \tau^{-3/2} \right)^e \left( -\tau \right)^v \left( \tau \right)^f = \tau^{-3 f / 2}. \end{equation}
Now suppose we have a configuration with some black hexagons.  Its weight is just that of the all white configuration multiplied by appropriate ratios of theta symbols, tetrahedron symbols, and quantum dimensions.  More specifically, for each white hexagon that one turns into a black hexagon, one must multiply the edge contribution by $(\tau / \tau^{3/2})^{-6}$, the vertex contribution by $(\tau^{3/2} / (- \tau))^6$, and the face contribution by $\tau^{-1}$.  The product of these is $\tau^5$, so that the weight of each such coloring is just $\tau^{5 \# (\text{ black hexagons})}$ times the weight of the all-white coloring.  The sum over colorings now just yields the critical Hard Hexagon model, whose free energy per vertex was obtained by Baxter \cite{Baxter} (eqn. 10).  We thus obtain
\begin{equation} G_\tau = {\left( \tau^{-3/2} \kappa_c \right)}^f \end{equation}
where 
\begin{equation} \kappa_c = \left( \frac{27 (25+11\sqrt{5})}{250} \right)^{1/2}. \end{equation}  From (\ref{eqn:norm}) we surmise
\begin{equation} \chi(\tau+1) = G_\tau {\tau}^{f-5/4 v} = {\left( \tau^{-3} \kappa_c \right)}^f. \end{equation}
We note that this is sharper than Tutte's bound \cite{Tutte} of $\text{const. } \tau^{-f}$ on $\chi(\tau+1)$:
\begin{equation} 0.546^f < 0.618^f. \end{equation}
Tutte's bound is of course more general in that it applies to any net configuration on the sphere (whose regions are simply connected and don't border on themselves).  For completeness, we note from (\ref{eqn:tutte2}) and (\ref{eqn:fe}) that $c_0 = \tau^{-3/2} \kappa_c$.

\end{document}